# Direct Nano-Imaging of Light-Matter Interactions in Nanoscale Excitonic Emitters

Kiyoung Jo[1], Emanuele Marino[2], Jason Lynch[1], Zhiqiao Jiang[2], Natalie Gogotsi[3], Thomas P. Darlington[4], Mohammad Soroush[5], P. James Schuck[4], Nicholas J. Borys[5], Christopher Murray[2,3], Deep Jariwala[1]*

[1]Department of Electrical and Systems Engineering, University of Pennsylvania, PA, 19104, United States

[2]Department of Chemistry, University of Pennsylvania, PA, 19104, United States

[3]Department of Materials Science and Engineering, University of Pennsylvania, PA, 19104, United States

[4]Department of Mechanical Engineering, Columbia University, New York, New York, 10027, United States

[5]Departement of Physics, Montana State University, Bozeman, Montana, 59717, United States

*Corresponding author: dmj@seas.upenn.edu

## Abstract

Strong light-matter interactions in localized nano-emitters when placed near metallic mirrors have been widely reported via spectroscopic studies in the optical far-field. Here, we report a near-field nano-spectroscopic study of the localized nanoscale emitters on a flat Au substrate. We observe strong-coupling of the excitonic dipoles in quasi 2-dimensional CdSe/Cd$_x$ZnS$_{1-x}$S nanoplatelets with gap mode plasmons formed between the Au tip and substrate. We also observe directional propagation on the Au substrate of surface plasmon polaritons launched from the excitons of the nanoplatelets as wave-like fringe patterns in the near-field photoluminescence maps. These fringe patterns were confirmed via extensive electromagnetic wave simulations to be standing-waves formed between the tip and the emitter on the substrate plane. We further report that both light confinement and the in-plane emission can be engineered by tuning the surrounding dielectric environment of the nanoplatelets. Our results lead to renewed understanding of in-plane, near-field electromagnetic signal transduction from the localized nano-emitters with profound implications in nano and quantum photonics as well as resonant optoelectronics.

# Introduction

Understanding of light-matter interactions in materials with strongly resonant properties and deep-subwavelength dimensions is important for both basic science and nano-opto-electronic applications. In most cases, light does not modify the electronic dispersions of the materials. This is because either the material dimensions are much greater than the wavelength λ or the material does not have an electronic dipole resonance at λ resulting in weak coupling. When the coupling between light and dipoles in matter becomes stronger, the rapid exchange of energy between photon states and electric dipole resonances leads to the formation of part-light, part-matter quasiparticle states called polaritons.[1] Different types of dipoles form different types of polaritons, including plasmon-polaritons in metals, exciton-polaritons in semiconductors, and phonon-polaritons in dielectrics in the IR range.[2–5] Such strong light-matter couplings require a confinement of light in a low dimensional material or interface to efficiently interact with the dipoles. For the case of surface plasmon polaritons (SPPs), light is confined to a dielectric/metal interface and forms a propagating electromagnetic wave along the surface.[6] Conversely, the formation of exciton-polaritons requires an intrinsic optical resonance of the medium to overlap with trapped light wave-packets in an optical cavity medium such as Bragg mirror dielectric microcavity or plasmonic cavity.[1,5] Propagating modes of exciton-polaritons were also observed with help of scattering type nanoprobe[7,8] and microcavities[9,10].

In reduced dimensional materials, the exciton that is induced by light-excitation interacts strongly with the surrounding medium due to the lack of dielectric screening. The binding energy of the excitons in 2-dimensional (2D) $WSe_2$ (0.78 eV)[11], 1-dimensional (1D) single-walled carbon nanotubes (0.3-0.4 eV)[12] and 0-dimensional (0D) CdSe quantum dots (0.2-0.8 eV)[13] are significantly larger than room temperature thermal energy (0.025 eV). The large exciton binding energy of the nanomaterials makes excitons – not free carriers - the dominant excited species, resulting in stronger light-matter interaction. Strong light-matter coupling in excitonic nanomaterials has been investigated in many ways such as exciton-polaritons in 2D $MoS_2$ placed in an optical cavity[14], exciton-plasmon polaritons of 2D $WSe_2$[15], 0D CdSe/ZnS quantum dot placed in plasmonic cavities[16,17], and surface plasmon polaritons at 2D $MoS_2/Al_2O_3$/Au interfaces[18]. Most of these studies have been conducted in either diffraction-limited optical setups[14–16] or via non-optical excitation techniques such as electron energy loss spectroscopy [19]. However, less has been studied in terms of imaging strong light-matter coupling in nanoscale materials when excited in the near-field at optical frequencies.[17,20] Further, the impact of the nano-probe and complex nano-optical fields on dipole interactions as well as energy guiding and transduction at these deep sub-wavelength scales remains largely unexplored.

Tip-enhanced nano-spectroscopy has paved the way for direct nano-resolution spatio-spectral imaging of the emission of nanomaterials at optical frequencies.[1,21] By taking advantage of plasmonic gap mode confined in the nano-gap between the plasmonic tip and the substrate, this technique has enabled the visualization of optical responses from sub-wavelength semiconductor structures such as fluorescence/radiation patterns of quantum dots[20], strain-induced Raman and fluorescence shifts[22,23], lateral heterostructures of van der Waals semiconductors[24,25], and even the localized excitonic emission from nanobubbles in 2D semiconductors.[26] Most studies on the emission using tip-enhanced nano-spectroscopy are performed in contact mode to maximize the plasmonic gap mode confinement resulting in strong light-matter coupling along the normal direction to the surface. For example, strong light-matter interactions in CdSe/ZnS quantum dots using the plasmonic gap mode was achieved leading to exciton-plasmon polariton formation.[27] In addition, brightening of the dark exciton in monolayer transition metal dichalcogenides via the Purcell effect has also been observed.[28] Yet, investigations for inelastic emission or scattering with the tapping mode configuration have been limited.[29,30] Since the AFM tip oscillates in tapping mode, it is still close to the surface hence it can be considered a near-field signal. Recently, it was reported that tapping mode, tip-enhanced Raman maintains sub-wavelength resolution capability and is even beneficial in terms of a charging-free measurement tool by preventing hot-carrier injection.[31,32] However, the role of the tapping mode tip in emission is still elusive.

Herein, we report selective light-matter interactions of quasi-2D emitters on dielectric/Au or $SiO_2$/Si substrates with contact and tapping mode tip-enhanced spectroscopy. When using contact mode, we observe exciton-plasmon polaritonic emission. Conversely, when using tapping mode tip-enhanced spectroscopy, we visualize in-plane near-field radiation and radiative energy propagation via SPPs launched by the emitters on dielectric/Au or $SiO_2$/Si substrates. By placing the nanoscale emitters, such as CdSe-$Cd_xZnS_{1-x}$S nanoplatelets and $WSe_2$ nanobubbles, on dielectric/Au interfaces, we observe radiative fringe patterns that are indicative of sub-wavelength energy transfer from the nanoscale excitonic emitters to the plasmonic Au substrate. We further observe that dielectric permittivity and thickness are key parameters that control the observed fringe patterns and corresponding energy transfer. Our results facilitate a deeper understanding of near-field radiation from low-dimensional and hetero-dimensional excitonic systems.

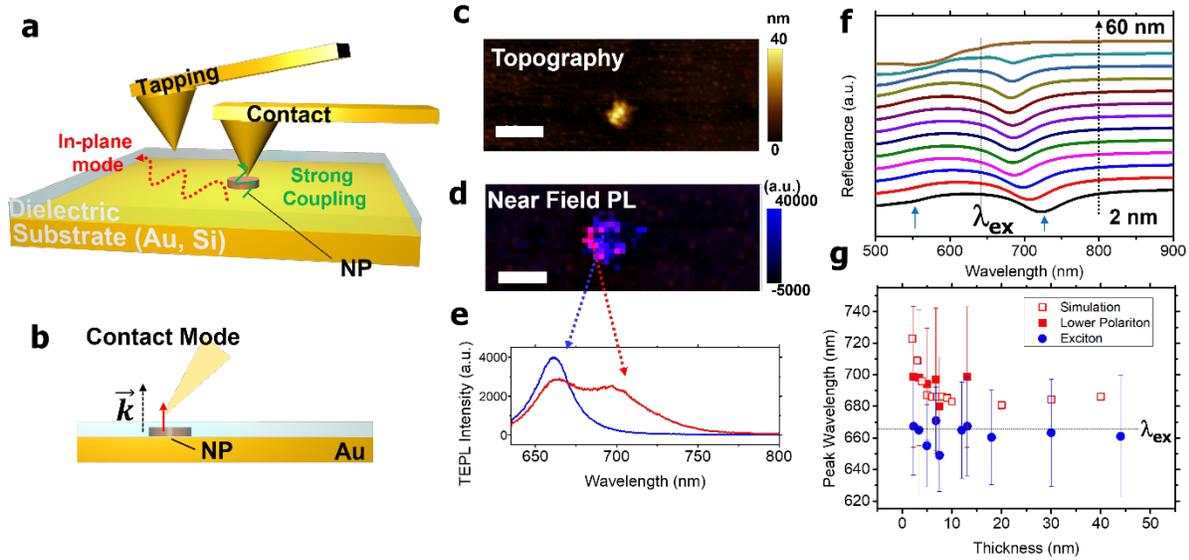

Figure 1. (a) Schematic representation of contact and tapping mode tip-enhanced scanning probe spectroscopy on dielectric/CdSe-Cd$_x$ZnS$_{1-x}$S nanoplatelet (NP)/Au system. (b) Schematic representation of the wavevector orientation of the light collected by the Au tip in the contact mode. (c) Topography image of a TiO$_2$ (5nm)/NP/ Au and (d) concurrent near field (contact-tapping mode) photoluminescence (PL) hyperspectral map and (e) the corresponding spectra of NPs with the thicknesses of 2.3 nm (red) and 38 nm (blue). Blue color at (d) represents TEPL intensity at 663 nm, while red indicates TEPL intensity at 700 nm. The near-field emission map and the spectra at (d) and (e) were obtained by subtracting the tapping mode TEPL from contact mode TEPL. (f) Simulated reflectance spectrum of TiO$_2$ (5 nm)/NP/Au near the tip. (g) Experimental excitonic (blue filled circle), lower polaritonic (red filled rectangle) and the simulated lower polaritonic (red empty rectangle) peak position of the lower polaritonic emission as a function of NP thickness. The error bars were taken from full width at half maximum (FWHM) of the spectrum. Scale bar = 0.5 μm.

*Strong coupling between NP excitons and tip plasmons in contact mode TEPL*

As a nanoscopic antenna, the plasmonic Au tip plays distinct roles in the collection of optical signals when working either in contact or tapping mode. Figure 1a describes the two different detection modes of tip-enhanced photoluminescence (TEPL). In tapping mode, the tip oscillates with a 20 nm amplitude and is only a few nanometers away from the surface at its extrema in the oscillation. For this reason, tapping mode operation does not result in strong light-matter confinement, precluding the formation of strongly coupled, hybridized light matter states such as exciton-polaritons. Conversely, in contact mode the tip-sample distance is < 1 nm. Subsequently, either plasmonic gap- or tip-mode triggers strong coupling between the excitons of the emitter and light depending on the existence of the plasmonic substrate.[33,34] Since the tip is very close to the sample in contact mode, the collection volume is tightly confined to the nanoscale cavity created by the tip and Au substrate. The majority of detected optical signals emerge from directly beneath the tip, and the dipolar orientation of the emission is normal to the substrate (Figure 1b). To probe this effect, we investigate strong light-matter coupling between the quasi-2D CdSe-$Cd_xZnS_{1-x}S$ core-shell nanoplatelets (NPs)[35] and the plasmonic tip collecting spectral data at each pixel with ~20 nm (tip radius) spatial resolution while concurrently mapping the topography. The NPs inherently have a 2D electronic band structure. Hence, the excitons in NPs are free to move in-plane, but they are also confined along the z-axis resulting in them having in-plane dipole moments.[36,37]

We synthesize well dispersed, highly luminescent CdSe/$Cd_xZnS_{1-x}S$ core/shell NPs by following literature.[38–40] Using TEM, we measure the size of the NPs to be 40.2 ± 2.9 nm (length) × 16.1 ± 1.7 nm (width) × 2.8 ± 0.5 nm (thickness). The samples were prepared by spin-coating a highly dilute (0.001 mg/mL) NP dispersion on an Au substrate followed by the deposition of a 5 nm dielectric layer, either $TiO_2$ or $Al_2O_3$, using atomic layer deposition. Topography images show that the NPs are aggregated to form 200 nm-wide clusters, which are likely resulting from the high dilution factor (Figure 1c and S3a). The thickness of the cluster varies between 2.3 and 50 nm, and the emission spectra varies drastically as a function of the cluster thickness (Figure 1e, g). This local variation of the TEPL spectrum is evident in Figure 1d. For a 40 nm thick NP cluster, a single emission at 664 nm was observed reflecting the excitonic emission of the NP (Figures 1d, 1e and S1). On the other hand, a 2.3 nm thick cluster, corresponding to the thickness of a single NP showed an additional feature at 699 nm (red curve in Figure 1e). This feature is not observed from NPs dispersed in solution (Figure S1), and we attribute it to the lower exciton-polaritonic emissions, respectively. To prove the 699 nm emission is attributed to the strong light-matter interaction, we simulate the reflectance of the $TiO_2$ (5 nm)-NP on Au in the vicinity of an Au tip with side illumination of 65° matching our experimental setup (Figure 1f, S2). The two polaritonic peaks (721 nm, 552 nm marked by blue arrow at Figure 1f) were observed when the NP thickness was 2 nm and the peak splitting narrowed down as the NP thickness increased. We attribute this peak splitting to the enhanced electric field of the tip-mode plasmon which mainly couples with the top layers of the NP. The 2D E-field intensity map indicates that the tip-mode induced electric field passes through several NPs, but the field intensity decreases as the thickness increases (Figure S2). This is further confirmed by the fact that the peak splitting was observed up to 13 nm thick NP aggregates which is far beyond the length scale of confinement of gap plasmon mode.[41] The simulated lower polariton peak matches the experimental values, suggesting that strong coupling can be induced by the plasmonic tip and side illumination (Figure 1g).

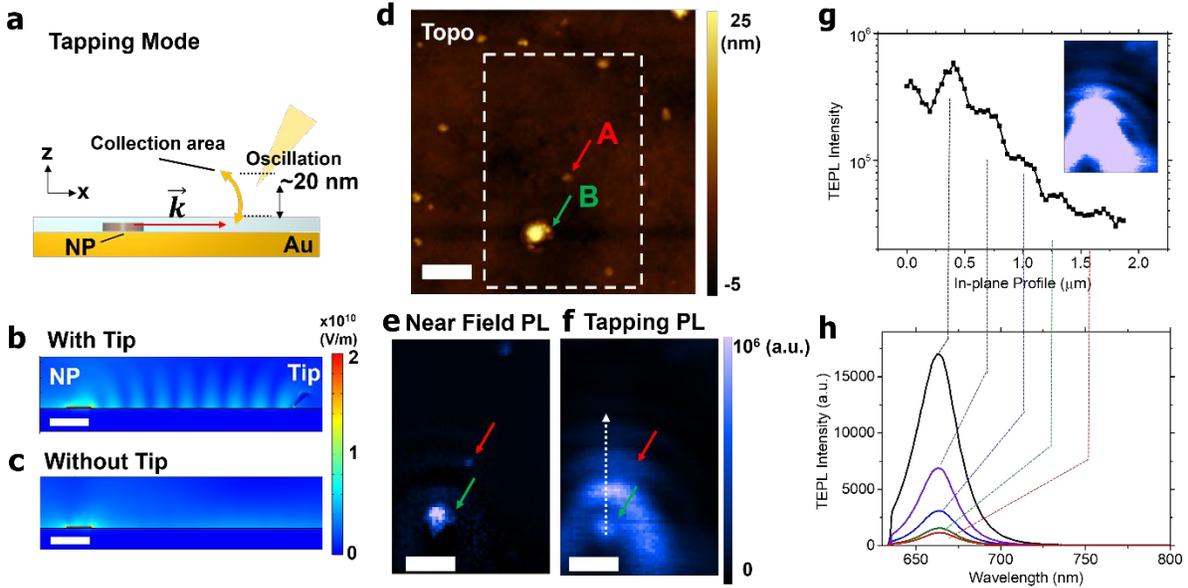

Figure 2. (a) Schematic representation of the wavevector orientation of the light collected by the Au tip working in tapping mode for a dielectric-NP on Au. (b, c) 2D map of the simulated E-field strength for $Al_2O_3$ (5 nm)/NP/Au at 664 nm wavelength (b) with and (c) without the Au tip. (d) Topography image of $Al_2O_3$ (5 nm)/NP/Au and (e) corresponding hyperspectral photoluminescence map of the dashed region of (d) with near-field TEPL (contact mode TEPL subtracted by tapping mode TEPL) and (f) tapping mode TEPL at 664 nm emission. A 633 nm laser was used for excitation. (g) Photoluminescence fringe intensity profile along the dotted arrow line in (f). Inset: tapping mode TEPL map in (f) with a saturated intensity scale. (h) the corresponding photoluminescence spectra. Scale bars indicate 0.5 μm. The red and green arrows represent the NPs with face-down (red) and edge-up (green) configuration.

*In-plane interference of NP excitonic emission at a dielectric/Au interface via tapping mode*

In tapping mode, the tip is a few nanometers away from the substrate as it oscillates. The tip can interact with a larger area of the sample than contact mode; since the Au tip serves as a scatterer in the near-field, this allows for the collection of light propagating along a dielectric/metal interface (Figure 2a). At the dielectric/metal interface, light is confined to the interface and propagates along the surface by forming surface plasmon polaritons (SPPs). SPPs can be reflected when they encounter either a microstructure or a change in dielectric permittivity at the surface.[42] In the dielectric/NP/Au system that we are investigating, NPs that are hundreds of microns away from the tip can be excited by a laser or by a SPP launched by the tip toward the NPs as reported in various studies using a scattering-type tip[43–46]. The emission from the NPs launch successive SPPs along the high index dielectric/metal interface and directly interferes with itself upon reflection from a distant Au tip. This observation is consistent with previous reports on SPPs launched by nano-emitters in the vicinity of the plasmonic graphene sheets.[47,48] For the SPP excitation from NPs, we speculate that the SPP can be coherently excited by excitonic photoluminescence because exciton coherence dephasing time of the NPs at room temperature (~10-25 fs)[49] is comparable to the dephasing time of the SPP at $Al_2O_3$/Au interface (~10 fs)[50]. In tapping mode, the Au tip not only acts as the antenna that helps collect near-field, in-plane electromagnetic waves, but it also as a reflector for SPPs because of its proximity to the dielectric/metal interface. As a result, the tip reflects the SPPs back to the emitter forming a standing wave. Similar standing waves have been observed in different types of polaritons via scattering mode n-SOM i.e. near-IR SPP in graphene[43,44,51], exciton-polaritons in bulk $WSe_2$,[45] and phonon-polaritons in hBN metasurfaces[2,46,52]. In our work, both NPs and the tip can launch the SPPs along the dielectric/Au interface. To evaluate the contribution of each case, we simulated cross-sectional E-field profile of dielectric/NP/Au in the vicinity of the Au tip engaged in the tapping mode, i.e. 20 nm away from the Au surface, to visualize the generation of these standing waves. Figure 2b and 2c verifies that the fringes only appear when both the NP and tip are present and close to the Au substrate which supports our explanation that the tip reflects the SPP and forms fringe patterns. As the tip-NP distance increases, the E-field intensity periodically changes, proving the standing wave condition changes with tip-NP distance (Figure S11). Moreover, a comparison of the $E_z$ field map with and without the tip demonstrates that surface-confined electric field persists from SPPs regardless of the presence of tip (Figure S4a, S4b). This is also consistent with our explanation that the excitonic emitter launches SPPs as previously reported in graphene plasmon induced by nanoemitters.[47,48] Yet, the tip could be either a reflector of SPPs launched by NPs or a SPP source launched by incident light. We evaluated the contribution of the tip plasmon toward the formation of the fringes by simulating the system without a NP (Figure S12b). In this simulation, we placed a dipole far away from the Au substrate, but the tip is kept close to the substrate. We found that the tip launches SPPs (Figure S12b - d) though its strength is 20-100 times smaller than the fringes found in NP-tip system. We note that the quantitative analysis may not be valid, and the tip-induced SPP is comparable to SPP launched by NP when the nanoemitter is not highly emissive i.e. off-resonant excitation, low quantum yield nanoemitter. In nearly on-resonant excitation to exciton wavelength, it is hard to distinguish SPPs launched by the NP vs the tip. Therefore, we can generalize the mechanism of fringe formation as one achieved by interference between SPPs launched by tip and NP interferes.

With this mechanism, a standing wave is expected to form between the NP and the Au tip with the following condition,

$$\frac{\lambda_{SPP}}{2} N = L_n$$

where λ$_{SPP}$ is the wavelength of SPP, *N* is an integer and *L*$_n$ is the length of cavity which is determined by the distance between the tip and the emitter.

Figure 2b shows the multiple fringes emerging between the NP and the Au tip which we refer to as a SPP standing wave. The Fourier transform of the fringe pattern yields a period of 317 nm, which is consistent with the expected standing wave SPP period $\frac{\lambda_{SPP}}{2}$= 320 nm. The expected period is calculated according to the dispersion relation of SPPs,

$$\lambda_{SPP} = \lambda_o \sqrt{\frac{\epsilon_d+\epsilon_m}{\epsilon_d\epsilon_m}}$$

where λ$_{SPP}$ and λ$_o$ are the wavelengths of SPP and PL emission, respectively and $\epsilon_d$ and $\epsilon_m$ are dielectric constants of the dielectric and metal, respectively.

The simulations presented in Figure 2b indicate that the Au tip acts as a reflector for SPPs, and that the standing wave condition can be controlled by changing the distance between the NP and Au tip. As the tip scans the surface near the NPs, the interference between the SPP launched by the excitons in the NP and reflected SPPs forms fringe patterns of the emitted radiation. To verify the interference pattern, we obtained hyperspectral TEPL maps with ~20 nm spatial resolution of an Al$_2$O$_3$ (5 nm)/NP/Au system. The topography and corresponding hyperspectral TEPL maps were obtained with contact and tapping mode during the same scan (Figures 2d, 2e, and 2f) and can therefore be used for fair comparison. As stated, NPs are quasi-2D nanocrystals which support excitons with dipoles oriented along the crystal plane. However, due to their relatively small aspect ratio (NP width/thickness ~14) compared to 2D materials (ideally infinite), either edge or plane of NP are in contact to the substrate which leads to different dipole orientation with respect to the substrate (Figure S1b). It has been reported that the transition dipole orientation of edge-up assembled NP (edge of NP is in contact with substrate) is perpendicular whereas the dipole orientation of face-down assembled counterparts (NP plane is in contact with substrate) is parallel with respect to the substrate.[37] The NP clusters A and B in Figure 2d represent NPs with face-down (A) and edge-up (B) configuration (Figure S1c). Considering the size of a single NP plate (40×16×2.8 nm$^3$), region A was determined to be a face-down configuration by its thickness (12 nm) which corresponds to the thickness of 2 NPs surrounded by 2 nm oleate ligand layer. Cluster B was characterized as edge-up configuration with 2 NPs with ligands which coincides with 40 nm thickness. Near-field TEPL of the two different regions shows spatially localized emission at 664 nm (Figure 2e). In contrast, a boomerang-shaped fringe pattern arises from the NP at spot B at 664 nm when measured in tapping mode (Figures 2f and 2h) while the NP at spot A does not. This source-selective fringe pattern can be explained by the different transition dipole orientation of A and B which we will explain below. The comparison between TEPL under different modes demonstrates that the tapping mode detects interference caused by in-plane emission from the NP while the contact mode does not. Fourier analysis of the fringe pattern (see SI for details, Figure S5) reveals that the fringe period in the boomerang shaped pattern was 322±119 nm at the symmetry axis of the boomerang (arrow line in Figure 2f) which is similar to the simulated fringe period (Figure 2b, g). Note that the error is measured using FWHM of the peak appearing in the Fourier transform. The large error bars are due to the lossy plasmonic cavity of dielectric/metal medium. The similarity in fringe periods between the experiment and the simulation suggests that SPP standing waves are responsible for producing the boomerang shaped fringe patterns in near-field emission maps. SPPs in the near-field are usually measured in scattering mode using dark-field imaging to collect the scattered light that has propagation directions that are highly misaligned from the normal direction.[53] Here, scattered SPPs are measured in emission mode since the SPPs are launched from the excitonic emission of highly emissive NPs and the tip is close enough to the interface to allow collection of the scattered light from the interface. As the tip scans the sample close to an excitonic emitter, it encounters standing wave conditions at specific locations and collects all the scattered light by the tapping mode tip. This indicates that the tapping mode maintains sub-wavelength resolution. Hence, the tapping mode TEPL map is not just a far-field spectrum, but it is the direct imaging tool for in-plane light propagation. Further, the fringes in emission mode are valuable in investigating the contributions of different exciton dipole orientations on the excitation of SPP which we cover in the next section. The boomerang shaped fringes appeared until the 5th fringe ~ 1.7 μm away from the source with the TEPL intensity diminishing below detection levels beyond that, as shown in Figure 2g and 2h. The TEPL intensity at the antinode was fitted with an exponential decay curve, revealing a decay constant of 324 ± 0.01 nm (Figure S6a). Since the tip was away from the NP cluster, the interference intensity decreases due to the lossy cavity of the dielectric/Au interface.

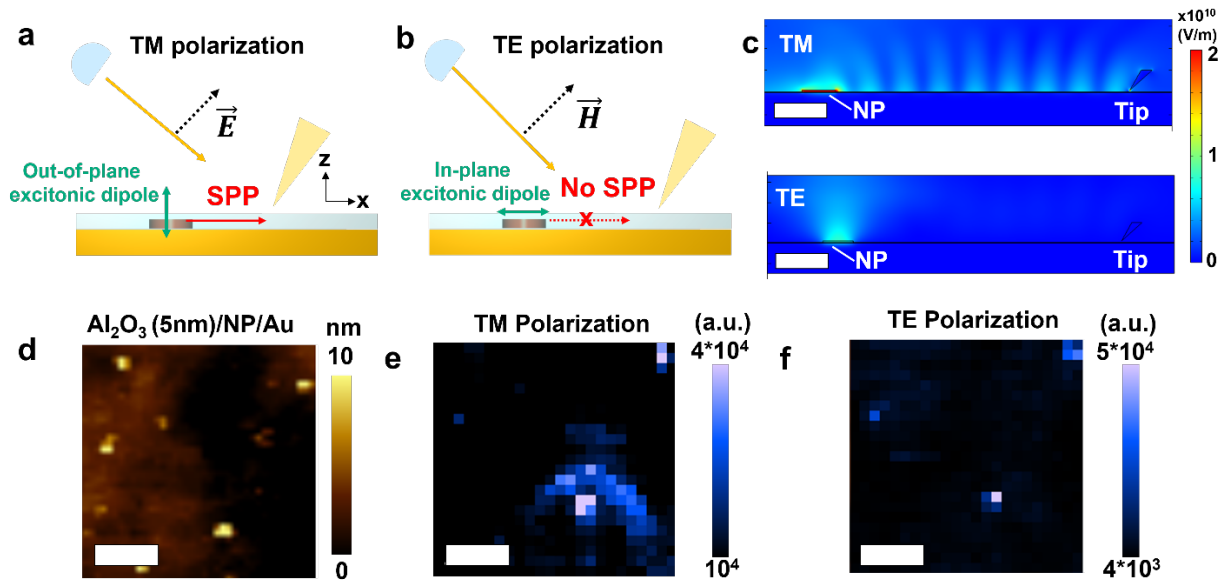

Figure 3. Schematic representation of the dielectric/NP/Au system with Au tip under (a) TM polarization and (b) TE polarization. (c) Simulated identical system under TM polarization (top) and TE polarization (bottom). (d) AFM topography image of Al$_2$O$_3$ (5 nm)/NP/Au and hyperspectral tapping mode TEPL under (e) TM polarization and (f) TE polarization. Scale bar indicates 0.5 μm.

*Polarization dependent standing wave surface plasmon polaritons*

The coupling between the emitters and the SPPs depends on the polarization of the fields of the emitted light and the excitation laser. For the excitonic emission, the directionality of the emitted light is governed by the orientation of the excitonic transition dipole. For example, the bright exciton of 2D semiconductors (i.e. WSe$_2$, MoSe$_2$) has an in-plane transition dipole, and therefore, maximum emission is collected by a detector that is oriented perpendicular to the basal plane of the 2D semiconductor. Similarly, for quasi-2D CdSe/Cd$_x$ZnS$_{1-x}$S NPs, the degenerate transition dipole is oriented along the plane of the NP.[36,54,55] However, as stated above, edge-up and face-down assembled NPs have different excitonic dipole orientation which makes a comparison of PL maps of the two different types of NP clusters more valuable in investigating dipole-orientation dependent SPP formation (Figure S1b). In the as-prepared sample, the orientation of NP clusters as well as its transition dipole orientation can be either perpendicular or parallel depending on the assembled configuration of individual NP clusters. Therefore, the individual NP clusters excite excitons with different dipole orientation depending on the laser polarization. For example, the edge-up NP cluster dominantly forms excitons with out-of-plane dipole orientation with respect to the substrate when the cluster is excited by the transverse magnetic field (TM) of incoming light. Because the exciton dipole orientation is parallel to the field orientation of surface plasmon in this case, it results in strong coupling, and therefore, the edge-up assembled NPs can launch the SPP (Figure 3a).[37] On the other hand, the face-down assembled NP clusters forms dominant in-plane oriented excitons with transverse electric field (TE). Since exciton transition dipole orientation is perpendicular to field direction of surface plasmon, it can neither couple with the plasmon nor launch the SPP (Figure 3b).

Considering dominant dipole orientation of NP clusters and polarization of excitation laser, only combination of TM polarized light and the edge-up assembled NP cluster can launch SPPs. In the dielectric/NP/Au system under TE polarized light, we do not expect any standing wave patterns regardless of the types of assembly configuration of NPs (edge-up and face-down). For edge-up NPs, out-of-plane exciton dipole is not the dominant orientation under TE polarized excitation laser. For face-down NPs, TE polarized excitation light (633 nm) excites predominantly in-plane excitonic dipoles in NPs leading to TE polarized emission that does not couple with the surface plasmons. Our finite element simulations support our explanation as shown in Figure 3c. TM polarized emission from NPs forms fringes at the lateral cavity between the NP and the tip whereas TE polarized emission does not. To verify the simulations with experiments, we conducted a polarization-dependent TEPL map to evaluate the relationship between the excitonic dipole and fringe formation (Figures 3d, 3e, and 3f). The tapping mode TEPL response differs with respect to the polarization of the excitation laser. Fringe patterns are observed for TM polarization while the patterns disappear under TE polarization agreeing with our hypothesis that the out-of-plane transition dipoles excited by the TM polarized laser emits in-plane directional light, successively launching an SPP (Figures 3a and 3e). On the other hand, TE polarized light excites excitons with in-plane transition dipoles, leading to negligible in-plane SPP generation (Figures 3b and 3f). We further note that fringe patterns only appear for thick NP clusters (>16 nm) whereas they do not appear for thin NP clusters (~3 nm). This observation again coincides with our claims that the edge-up NP clusters launch SPP while the other do not.

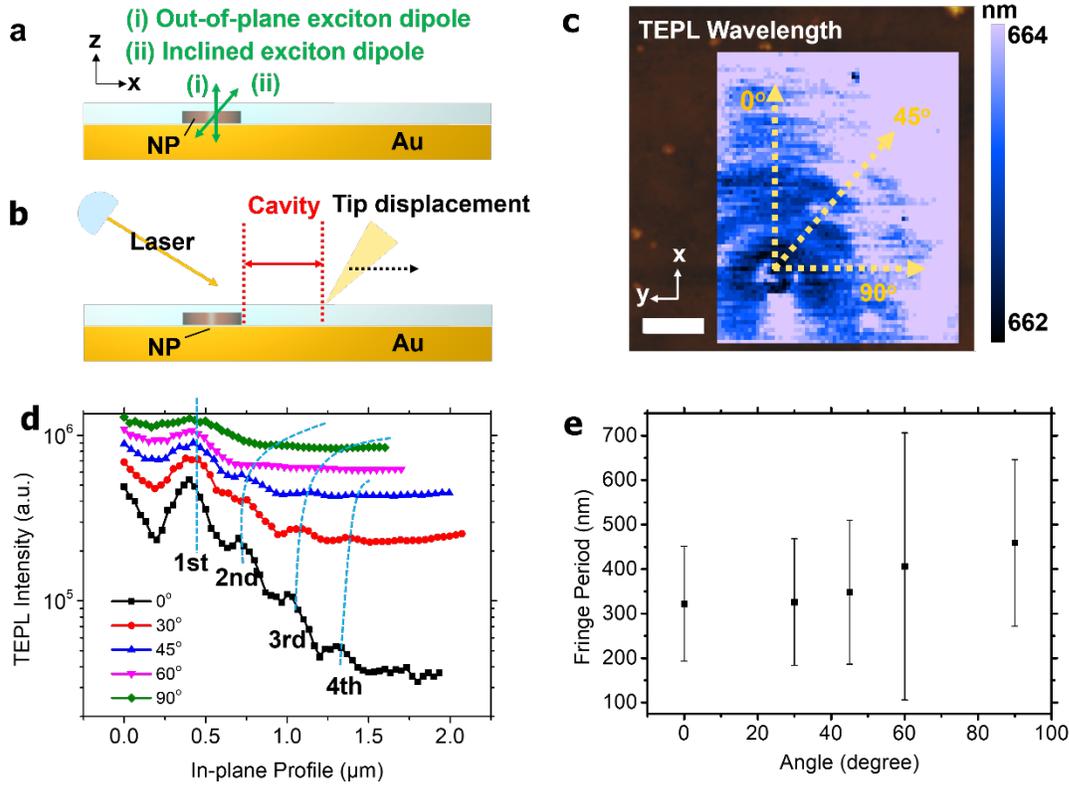

Figure 4. (a) Schematic representation of exciton transition dipole orientation (i) out-of-plane and (ii) inclined dipole orientation (b) Schematic representation of NP-tip lateral cavity length and tip displacement. (c) Tapping mode TEPL wavelength map of $Al_2O_3$ (5 nm)/NP/Au and (d) TEPL intensity line cuts at different direction angles. (e) Fringe period as a function of angle. Scale bar indicates 0.5 μm.

*Fringe shape and directionality of the generated surface plasmon polariton*

If the transition dipoles of the NPs are oriented normal to the substrate (Figure 4a), the fringe pattern would be circular because the SPPs would be launched uniformly in all in-plane directions.

However, the fringe pattern observed in the tapping mode TEPL map (Figure 2f) has a parabolic shape which reveals that the transition dipole orientation in the NPs is not normal but inclined. This claim was supported by the 3D finite element simulation that an out-of-plane dipole (Figure S9c) launches circular fringes while an inclined dipole (Figure S9d) creates parabolic or boomerang-shaped fringes due to the broken symmetry. The dependence of the shape of the fringe pattern on exciton dipole is significant since it allows experimental probing of dipole orientations in nanostructured emitters which is non-trivial. Further, fringe patterns appear only at the cavity between NPs and tip (Figure 4b, S9c and S9d). The fringe disappears outside the cavity revealing the directionality of the fringes is determined by the tip displacement. Therefore, the transition dipole orientation and tip displacement during the scan matters in determination of the shape of fringes. To experimentally support this, we show PL wavelength map of identical $Al_2O_3$/NP/Au system (Figure 4c). All the fringes have similar peak wavelengths that are close to the excitonic emission wavelength of the NP. Note that the background exhibits an artificially high wavelength due to the small signal-to-noise ratio. There is negligible signal detected at non-excitonic wavelengths (Figure S3c). We therefore ensure that the PL from the NPs is mainly responsible for the fringe shape (Figure 4c). The boomerang-shaped fringe indicates that the dipole orientation is not perfectly out-of-plane but inclined based on our simulations (Figure S9c and S9d). The TEPL intensity profiles along lines oriented from 0 to 90 degrees with respect to the symmetry axis were investigated to understand the directionality of the NP PL emission (Figures 4d and 4e). The fringe period is observed to increase with increasing angle, eventually disappearing. Quantitatively, the fringe period is 322 nm at 0 degrees whereas it is 459 nm at 90 degrees. This difference suggests that interference does not occur at a high angle due to lack of light directly heading to the tip and therefore the NP-tip cavity does not form.

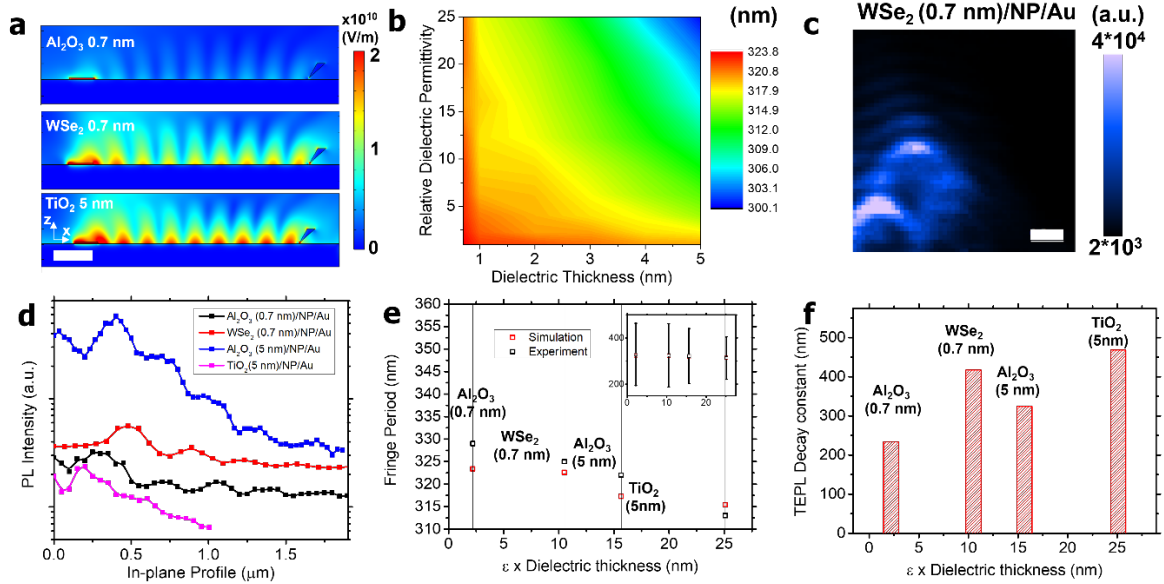

Figure 5. (a) 2D map of the simulated E-field intensity (4500 nm width × 600 nm height) with different dielectric layers. (b) Simulated fringe period as a function of relative dielectric permittivity ($\varepsilon$) and thickness of the dielectric layer. (c) Hyperspectral tapping mode TEPL map of WSe$_2$ (0.7 nm) / NP/Au at 664 nm. (d) TEPL intensity line cuts along with propagation direction. (e) Comparison of the fringe period between simulation and experiment. (f) TEPL intensity decay constants from exponential decay fit. Scale bars at (a) and (c) indicate 0.5 μm.

*Dielectric effect on the fringe period*

Since a SPP is a surface confined electromagnetic mode, it is subject to perturbations and alterations by changes in the dielectric medium at the interface. It is possible to observe these effects as small perturbations of the fringe pattern as the tapping mode TEPL reports the intensity distribution of the near-field electromagnetic wave bound to the dielectric/metal interface.

Therefore, we investigate the variation of the fringe pattern as a function of the permittivity and thickness of the dielectric. Simulations were performed using different dielectric materials and thicknesses. We used dielectrics that are readily deposited in thin uniform films via ALD or mechanical stamping such as $Al_2O_3$ ($\varepsilon = 3.13$ at 630 nm), $TiO_2$ ($\varepsilon = 5.02$ at 630 nm) and monolayer $WSe_2$ ($\varepsilon = 15$ at 630 nm). A 0.7 nm thick $Al_2O_3$ sample was prepared to compare with monolayer $WSe_2$. Figure 5a shows the simulated standing wave fringe patterns with different dielectric configurations. These results demonstrate that a larger difference in $\varepsilon$ between the dielectric material and metal creates a stronger confinement of electromagnetic waves at the interface. To generalize the dielectric effects on fringe periods, we simulated fringe patterns as a function of dielectric constant and thickness that is summarized in the 2D plot in Figure 5b. The period of the simulated standing wave increases with decreasing dielectric permittivity and dielectric thickness. We deposited $Al_2O_3$, $TiO_2$ and monolayer $WSe_2$ on as-prepared NPs on the Au substrate to verify experimentally the validity of the simulations. Note that the thickness of the NP cluster is ~30 nm for all samples and the emission spectrum peaks at ~664 nm (i.e. the excitonic wavelength) for all samples. Similar boomerang patterns were observed in $Al_2O_3$ (0.7 nm), $WSe_2$ (0.7 nm) and $TiO_2$ (5 nm)/NP/Au samples (Figure 5c, S7a, and S7b). However, distinct differences were observed in their periods (Figure 5d). This observation shows that even the thinnest dielectric confinement can modify the SPP propagation which can be recorded via the interference effect using a reflective tip. Compared to the fringe period (322 nm) of $Al_2O_3$ (5 nm)/NP/Au system, we observe 329±135 nm for $Al_2O_3$ (0.7 nm), 325±119 nm for $WSe_2$ (0.7 nm) and 313±92 nm for $TiO_2$ (5 nm)/NP/Au, respectively. Once again, the large error bars are due to lossy plasmonic propagation. The fringe period was plotted as a function of the product of dielectric permittivity ($\varepsilon$) and thickness (Figure 5e). The x-axis is selected to correlate the fringe periods to dielectric cavity volumes. The experimental values are in close agreement with the simulations. The fringe period increases as the dielectric permittivity or thickness decreases, which is in accordance with the simulation result in Figure 5b. Likewise, the TEPL decay constant, which directly correlates to energy transfer, shows the opposite relation with the dielectric constant of the medium i.e. larger $\varepsilon$ results in longer decay constant. Quantitatively, high dielectric materials such as $TiO_2$ and $WSe_2$ showed decay constants of 468±0.16 nm and 417±0.12 nm respectively (Figure 5f). This result coincides with our simulations (Figure 5a) showing that high dielectric medium results in high E-field strength of the SPP due to stronger confinement. The comparison between two different thickness of $Al_2O_3$ layers (324±0.11 nm for 5 nm $Al_2O_3$ and 234±0.12 nm for 0.7 nm $Al_2O_3$) shows that the thicker dielectric medium is more effective for in-plane light propagation (Figure S6).

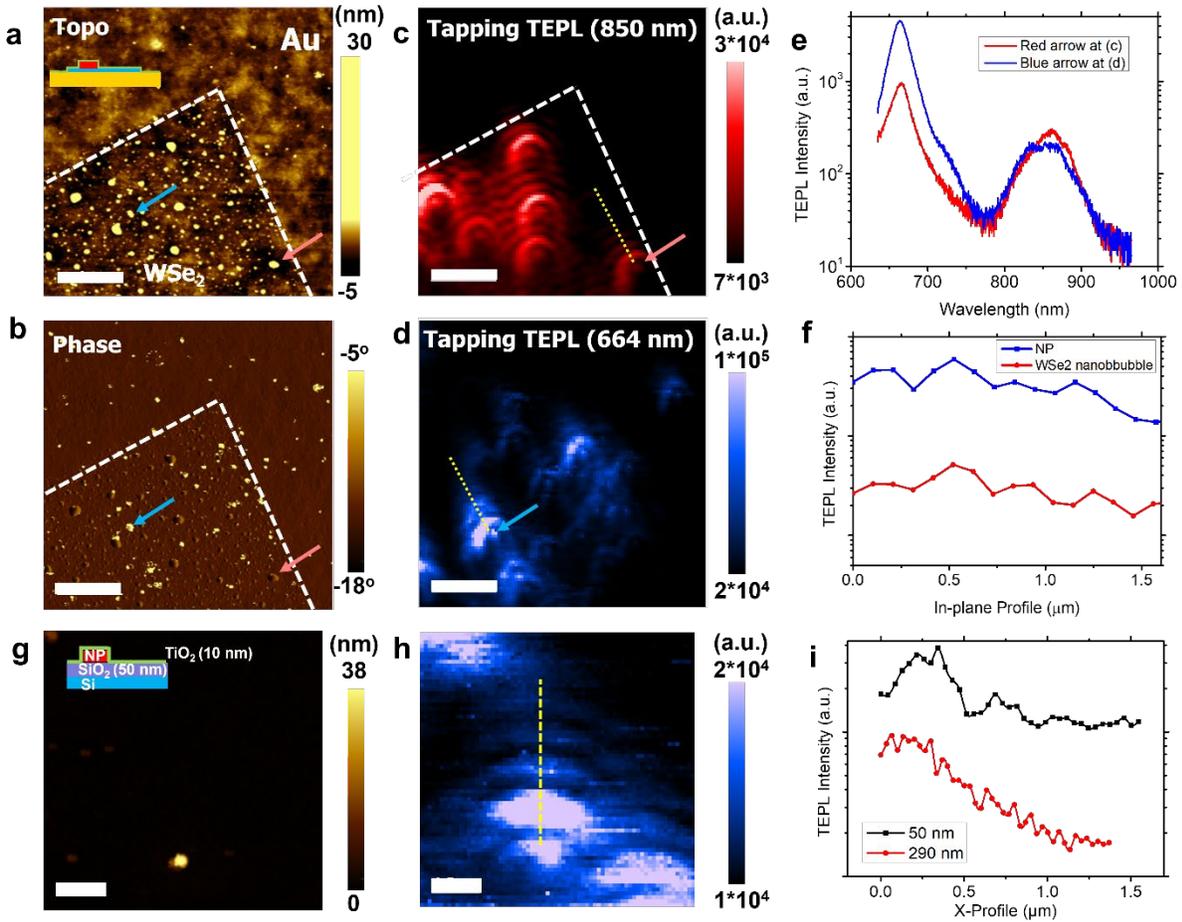

Figure 6. (a) Topography and (b) Phase map of Al$_2$O$_3$(0.7 nm)/NP/WSe$_2$/Au (c, d) Hyperspectral mapping of the tapping mode TEPL at (c) WSe$_2$ emission wavelength and (d) NP emission wavelength. (e) TEPL spectrum at red arrow at (c) (WSe$_2$ bubble) and blue arrow at (d) (NP). (f) TEPL intensity line cuts along the yellow dotted line at (c) and (d). (g) Topography of TiO$_2$(10 nm)/NP/SiO$_2$(50 nm)/Si and (h) hyperspectral tapping mode TEPL map of NP emission range (i) TEPL intensity line cuts along yellow dotted line in (h). The black curve represents 50 nm SiO$_2$ layer and the red represents 290 nm SiO$_2$. Scale bars at (a, b, c, d) indicate 2 μm. Scale bars at (g, h) indicate 0.5 μm. All measurements are conducted with 633 nm excitation laser.

*Universal emergence of fringe pattern by varying emitter or interface property*

To generalize this fringe phenomenon to other localized emitters, we explore the same effect in nanobubbles of monolayer $WSe_2$. Monolayer $WSe_2$ was prepared using mechanical exfoliation on a template-stripped Au substrate. To directly compare the fringe patterns produced by NPs to those of nanobubbles, we spin-coated NPs onto the exfoliated $WSe_2$ and then deposited a thin film of $Al_2O_3$. The ALD of the $Al_2O_3$ film led to an ensemble of nanobubbles with emission centered at 850 nm. The large redshift of the excitonic emission can be attributed to strain-induced bandgap lowering.[26] The monolayer $WSe_2$ flake contained nanobubbles that spanned 100 nm in diameter (Figure 6a). The phase map clearly distinguishes the NPs from nanobubbles (Figure 6b). NPs show an abrupt contrast change (blue arrow) whereas $WSe_2$ nanobubbles do not (red arrow). $WSe_2$ bubbles also act as the localized emitters which emit at 850 nm and launch fringes of 357±163 nm period (Figure 6c). Interestingly, fringe period increases to 414±136 nm when it is excited by 785 nm excitation laser (Figure S13c). This represents the contribution of tip-induced SPP excited by the laser. That is, with 785 nm laser which is close to excitonic emission of $WSe_2$ nanobubble, the tip-induced SPP wavelength is similar to that from $WSe_2$ nanobubbles. However, with high energy excitation lasers such as 594 and 633 nm, the tip launches SPPs with shorter wavelength while the contribution of SPPs launched by $WSe_2$ nanobubbles become smaller as it is excited by off-resonant laser and emits small PL intensity (Figure S13d). Considering that our generalized fringe mechanism is the interference between SPPs from excitonic emitter and tip, the fringes period shortens as the emitter – excitation laser energy difference becomes large. It indicates that the SPP launched by tip must be considered when the excitation laser is significantly different from the wavelength of excitonic emission at the nanoscale emitter. Meanwhile, the fringe period of NPs with 633nm excitation laser was 322±128 nm on the same substrate which is smaller than that of WSe$_2$ nanobubbles (Figure 6d). This is primarily because the wavelength of excitonic emission is more red-shifted for WSe$_2$ nanobubbles which requires a longer path length to attain the standing wave condition. Moreover, fringe shape of WSe$_2$ nanobubbles was circular which means its transition dipole orientation is nearly normal to the surface. On the other hand, the boomerang shaped fringes are launched by the NP clusters due to inclined dipole. To further investigate the effect of NP-tip cavity to form fringes, we rotate the sample and measured TEPL again (Figure S10). We observed same directional fringe formation after 90 degree sample rotation that again verifies that the standing wave is originated from NP - tip lateral cavity. In addition, the nanobubbles of different 2D semiconductor, WS$_2$, also launched fringes due to the SPP at 660 nm (Figure S14). This experiment and observation confirm that the localized excitonic emitters are all capable of launching SPPs on plasmonic substrates at different optical frequencies which can be observed in near-field in the form of fringe patterns with varying periods.

Finally, we also verify that the in-plane optical propagation of excitonic emission by localized emitters can lead to the formation of fringe patterns even in the case of non-plasmonic (dielectric) substrates. In this case the emitted light is expected to be confined into a waveguide mode that propagates and is reflected back from Au tip forming fringe patterns representing standing waves. To investigate this in-plane guided-mode light propagation in a dielectric layer, we fabricated TiO$_2$ (10 nm)-NP structures on SiO$_2$ (50 nm)/Si substrates rather than plasmonic Au substrates (Figure 6g). Note that TiO$_2$ is pre-selected for a higher refractive index (2.24) than that of the SiO$_2$ (1.46) to ensure larger light confinement in the SiO$_2$ layer. We observed a similar fringe pattern of the excitonic emission at 663 nm with a 50 nm thick SiO$_2$ layer (Figure 6h and S8c-f). This supports our hypothesis that the fringe is created by any surface light-confinement mechanism including waveguided photons. Finite element simulations further showed that the fringe period (439 nm) matches with the experimental measurements (442 ± 230 nm) (Figure S8d). In summary, by varying either the excitonic emitter or substrate, we can observe formation of standing wave fringe patterns in the near-field. We therefore conclude that these fringe patterns are a universal near-field phenomenon when a nanoscale emitter is placed on a highly smooth and index mismatched substrate.

*Conclusions*

We report a comprehensive study on the near-field interaction of a plasmonic tip with localized nanoscale emitters by spatially imaging their emission patterns using tip-enhanced nano-spectroscopy,. Taking advantage of nanoscale spatial-resolution capability of the tip, sub-wavelength interference of in-plane propagating electromagnetic modes can be analyzed under tapping mode operation while strong-coupling between excitons in nanoscale chalcogenide emitters with incoming photons confined in plasmonic gap mode can be observed under contact mode on Au substrates. Hyperspectral maps clearly illustrate that photons from localized emitters can be emitted in-plane that can be visualized ~1.7 microns away from the emitting source by virtue of standing waves formation. The interference period and the corresponding signal decay rate is governed by dielectric layer thickness and permittivity which also dictates the fraction of photons radiated in plane and the degree of confinement. The exciton transition dipole orientation and presence of the tip determines shape and directionality of the standing wave. Strain induced formation of localized emitters in a 2D dielectric medium is favorable in terms of in-plane radiation coupling efficiency. Our work shows that near-field scanning probe microscopy with metallic tips is a useful tool in imaging and analysis of nanoscale excitonic emitters including their radiation patterns as well as dipole orientations. In addition, our work helps understand energy transfer mechanisms and dynamics of excited state phenomena in emitters at deep sub-wavelength scales with both spectral and spatial information. The technique and approach could therefore serve as a useful tool for imaging, identifying and manipulating dipoles of even quantum emitters opening new avenues in classical and quantum nanophotonics.

*Online Contents*

The supporting information is available in the online version of the paper.

Synthesis procedure of CdSe/Cd$_x$ZnS$_{1-x}$S nanoplatelets, TEM and emission spectrum of the as-prepared nanoplatelets, Preparation of dielectric-NP on Au samples, details of near-field measurements, determination of the fringe period, COMSOL Multiphysics simulations for reflectance spectra in contact mode, polarization dependent 2D field profile in the vicinity of the tip, estimation of TEPL decay rate, contact mode hyperspectral map of NP, hyperspectral map of dielectric-NP on SiO$_2$/Si, hyperspectral map of Al$_2$O$_3$ (0.7 nm) and TiO$_2$ (5 nm)-NP on Au, simulated 2D field strength map in XY plane, TEPL maps after 90° rotation of the sample are all described in detail in supporting information.

Corresponding Author

E-mail address: dmj@seas.upenn.edu

Author Contribution

K.J. and D.J. conceived idea and designed research. K.J. fabricated samples, conducted near-field measurement and analyzed the data. E.M., Z.J. and N.G. synthesized the nanoplatelets under supervision of C.B.M. T.P.D., M.S., P.J.S. and N.B. assisted in data interpretation and experiment design. K.J. and J.L. conducted COMSOL Multiphysics simulations. K.J. and D.J wrote the manuscript. All authors discussed the results and commented on the manuscript.

**Acknowledgements**

D.J. acknowledges primary support for this work by the U.S. Army Research Office under contract number W911NF-19-1-0109. K.J. acknowledges the support from the Vagelos Institute for Energy Science and Technology Graduate fellowship. The sample fabrication, assembly and characterization were carried out at the Singh Center for Nanotechnology at the University of Pennsylvania which is supported by the National Science Foundation (NSF) National Nanotechnology Coordinated Infrastructure Program grant NNCI-1542153. Authors also acknowledge support from University of Pennsylvania Materials Research Science and Engineering Center (MRSEC) (DMR-1720530),usage of MRSEC supported facilities, and the Office of Naval Research Multidisciplinary University Research Initiative Award ONR N00014-18-1-2497. N.J.B. and P.J.S. acknowledge support from the National Science Foundation through award NSF-2004437.

# Supporting Information

# Direct Nano-Imaging of Light-Matter Interactions in Nanoscale Excitonic Emitters

Kiyoung Jo[1], Emanuele Marino[2], Jason Lynch[1], Zhiqiao Jiang[2], Natalie Gogotsi[3], Thomas P. Darlington[4], Mohammad Soroush[5], P. James Schuck[4], Nicholas J. Borys[5], Christopher B. Murray[2,3], Deep Jariwala[1]*

[1]Department of Electrical and Systems Engineering, University of Pennsylvania, PA, 19104, United States

[2]Department of Chemistry, University of Pennsylvania, PA, 19104, United States

[3]Department of Materials Science and Engineering, University of Pennsylvania, PA, 19104, United States

[4]Department of Mechanical Engineering, Columbia University, New York, New York, 10027, United States

[5]Departement of Physics, Montana State University, Bozeman, Montana, 59717, United States

# Experimental

## Synthesis of CdSe/Cd$_x$ZnS$_{1-x}$S nanoplatelets

**Synthesis of CdSe nanoplatelets:** Cadmium myristate precursor is prepared by following the literature.[1] Colloidal, rectangular CdSe nanoplatelets with a thickness of 4.5 monolayers are synthesized following the literature[2] with slight modifications.[3, 4]

340 mg of finely-ground cadmium myristate and 28 mL 1-octadecene (technical grade, ODE) are added to a 100 mL three-necked round-bottom flask with a 1-inch octagonal stir bar. The central neck is connected to the Schlenk line through a 100 mL bump trap, one of the side necks is equipped with a thermocouple adapter and thermocouple, and the other one is fitted with a rubber stopper. With a heating mantle, the flask is degassed at 100 °C for 30 min. In the meanwhile, a dispersion of 0.15 M selenium in ODE is prepared by sonication for at least 20 min. After switching to the atmosphere of the flask to nitrogen, the temperature of the reaction is increased to 220°C. 2 mL of 0.15 M Se/ODE dispersion are quickly injected by using a 22 mL plastic syringe equipped with a 16 G needle. After 20 seconds, 120 mg of finely-group cadmium acetate are added to the flask by temporarily removing the stopper. The flask is carefully rocked to ensure that the cadmium acetate powder does not stick to the side walls of the flask. The reaction is kept at 220°C for 14 minutes and then rapidly cooled with a water bath. 12 mL of oleic acid (technical grade, OA) and 22 mL of hexane are added when the temperature reaches 160°C and 70°C, respectively.

**Washing procedure of CdSe nanoplatelets:** The nanoplatelets are washed by following a procedure reported in the literature with modifications.[5] The mixture is first centrifuged at 8586 g for 10 min. The precipitate is then redispersed in 10 mL of hexane. The suspension is left undisturbed for 1 h, and then centrifuged at 6574 g for 7 min. The precipitate is discarded as it contains undesired 3.5 monolayer nanoplatelets. The supernatant is retained and transferred to a new centrifuge tube. 10 mL of methyl acetate are added to the supernatant, followed by centrifugation at 5668 g for 10 min. 6 mL hexane is used to redisperse the precipitate. Measuring the optical absorption spectrum is useful to confirm the removal of the unwanted 3.5 monolayer nanoplatelets, which are characterized by a lowest-energy absorption peak at 462 nm, while the 4.5 monolayer nanoplatelets are characterized by a lowest-energy absorption peak at 512 nm. If 3.5 monolayer nanoplatelets are still present in the dispersion, they can be removed by titrating methyl acetate and centrifuging until all 3.5 monolayer nanoplatelets are successfully removed. The final dispersion is stored in a glass vial in the dark.

**Growth of $Cd_xZnS_{1-x}S$ shell:** Cadmium ($Cd(Ol)_2$) and zinc oleate ($Zn(Ol)_2$) are synthesized according to the literature.[5,6] The growth of $Cd_xZnS_{1-x}S$ shell on CdSe nanoplatelets is performed by following the literature[5] with minor modifications.

10 mL of ODE, 0.4 mL of OA, 90 mg of cadmium oleate, 167.5 mg of zinc oleate, and an amount of 4.5 monolayer CdSe nanoplatelets in hexane equivalent to a 1mL with an optical density of 120/cm at the lowest-energy absorption peak are added to a 100 mL three-necked round-bottom flask with a 1-inch octagonal stir bar. The central neck is connected to the Schlenk line through a 100 mL bump trap, one of the side necks is equipped with a thermocouple adapter and thermocouple, and the other one is fitted with a rubber stopper. The mixture is degassed for 35 min at room temperature and for 15 min at 80 °C. In the meanwhile, a solution of 83 μL of 1-octanethiol (OT) in 7 mL of degassed ODE and 2 mL of degassed OA is prepared in the glovebox and loaded in a plastic syringe. 2 mL of degassed oleylamine are added to a second plastic syringe. The two syringes are removed from the glovebox. Afterward, the atmosphere of the reaction flask is switched to nitrogen and the 2 mL of OAm are injected. Using a heating mantle, the temperature of the reaction flask is increased to 300 °C. At 165 °C, the solution of OT in ODE and OA is injected at a rate of 4.5 mL/h. After complete injection, the temperature of the reaction is maintained for an additional 40 min. The reaction mixture is cooled down to 240 °C by using an air gun, followed by using a water bath to cool to room temperature. At 40 °C, 5 mL of are added.

**Washing procedure of CdSe/Cd$_x$ZnS$_{1-x}$S nanoplatelets:** The reaction mixture is centrifuged at 6000 g for 6 min. The precipitate is redispersed in 5 mL of hexane while the supernatant is discarded. Methyl acetate is added to the dispersion until the mixture turned turbid, followed by centrifugation at 6000 g for 10 min. This process is repeated. The precipitate is redispersed in 3 mL of hexane and centrifuged at 6000 g for 7 min. The precipitate is discarded, containing aggregated nanoplatelets. The supernatant is retained and filtered through a 0.2 μm PVDF or PTFE syringe filter. The dispersion is stored in the dark under ambient conditions. The final dispersion is stored in a glass vial in the dark.

**TEM imaging:** For low-resolution TEM, a JEOL 1400 microscope was operated at 120 kV. For higher-resolution TEM, a JEOL F200 microscope was operated at 200 kV. During imaging, magnification, focus, and tilt angle were varied to yield information about the crystal structure and super structure of the particle systems. To prepare the dispersed nanocrystals for imaging, we drop cast 10 μL of a dilute (~0.1 mg/mL) dispersion of nanocrystals in hexane on a carbon-coated TEM grid (EMS). The grid was dried under vacuum for 1 hour prior to imaging.

**Spectrophotometry:** Absorption spectra of nanocrystal dispersions in toluene were measured by using a Cary 5000 UV-Vis-NIR spectrophotometer.

**PLQY:** PLQY measurements were performed by using the integrating sphere module of an Edinburgh FLS1000 Photoluminescence Spectrometer. The NCs were dispersed at a concentration corresponding to an absorbance of 0.1 at the excitation wavelength.

**Dielectric-NP on Au Sample preparation**

CdSe-CdS nanoplatelet was synthesized as reported.[7] The nanoplate solution (0.001 mg/ml in toluene) was spin-coated on the template-stripped Au substrate. Template stripped Au substrate was used for exceptionally low rms value (0.5 nm).[8] Dielectric layer ($Al_2O_3$, $TiO_2$) were deposited by atomic layer deposition (Cambridge Nanotech S200 ALD). Refractive index of dielectric layers was measured by Ellipsometer (Woollam VAS Ellipsometer).

**Tip-enhanced Photoluminescence Imaging**

LabRam-EVO Raman Spectrometer (Horiba Scientific) coupled with AFM setup (OmegaScope-R, AIST-NT) was used to conduct tip-enhanced photoluminescence measurement. After 633 nm laser was aligned to the apex of the tip, the sample is engaged with the feedback loop to measure the topography. Corresponding TEPL spectrum was obtained by both the contact and the tapping mode simultaneously. Each pixels in the hyperspectral map span 30*30 nm$^2$ and signals were collected for 100 ms. Near-field TEPL map and spectrum were extracted by subtracting the contact mode TEPL to the tapping mode TEPL.

**Determination of Fringe Period from Hyperspectral Images**

Discrete Fourier Transformation was performed to determine the fringe period. TEPL linecut profiles (Intensity vs *x*) were conducted by analyzing the data with to following equation.

$$F_n = \sum_{i=0}^{N-1} x_i e^{-\frac{2\pi j}{N} ni}$$

See figure S5 to find the FFT plots.

**Three-Dimensional Electromagnetic Wave Simulations.**

3D electromagnetic wave simulations were also performed in COMSPL specifically to determine the shape of the fringe patterns from the hyperspectral PL maps. To simply the simulations we introduced an Au slab at a certain distance away from the emitter which represents the superposition of multiple Au tips during a line scan (Figure S9a, b). This simulation visualized multiple fringes resulting from the standing wave formed by the in-plane cavity between the NP emitter and Au slab (figure S9c, d). Circular fringes appeared when the NP emitter dipole is perfectly out-of-plane (normal to Au surface) (figure S9c) while boomerang fringes were formed with the exciton dipole 45 degree-inclined in the XZ plane (figure S9d). This result shows that the excitonic dipole in NP in figure 4a is not perpendicular to the surface but inclined at a certain angle with respect to the surface. Moreover, the fringes were not observed outside the cavity which means the NP casts a shadow on the fringe patterns. In-plane cavity formation is therefore essential to detect the fringe patterns in a near field PL map. The fringe patterns in simulations only appear in the in-plane cavity between the NP and Au tip along the x-axis due to the fixed beam orientation during scanning. This is experimentally verified by acquiring successive TEPL measurements before and after sample rotation by 90º (figure S10). $Al_2O_3$/NP/$WSe_2$/Au samples were prepared to investigate fringe patterns of two different excitonic emissions at 664 nm from NP and 850 nm from $WSe_2$ nanobubbles. The directionality of the fringe propagation in the TEPL maps did not change after the sample rotation. The result indicates that interaction between directional emission of NP and Au tip is key part of collecting SPP fringes using this near-field scanning probe technique.

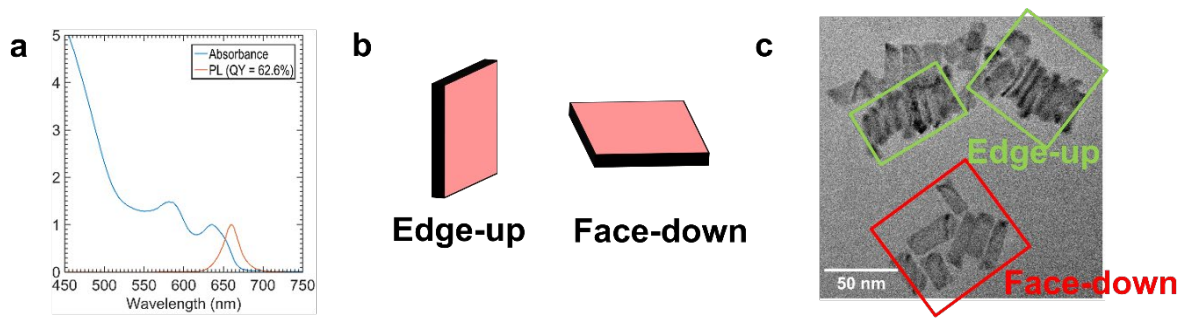

Figure S1. (a) Absorption and emission spectrum and PLQY of NP solution. (b)Schematic representation of edge-up assembled and face-down assembled NP cluster. (c) TEM image of NP. Region A (red) represents edge-up assembled NP while the region B (blue) refers to face-down assembled NP.

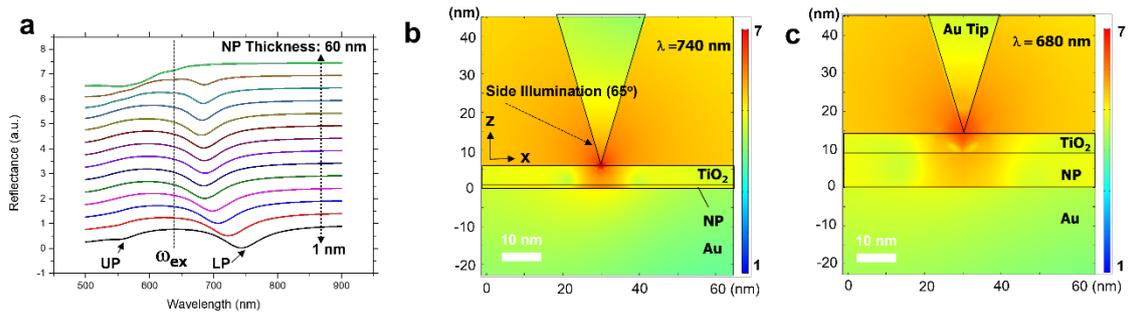

Figure S2. (a) Simulated reflectance spectra as a function of NP thickness under Au tip. Solid arrows indicate upper (UP) and lower polariton (LP) peaks, respectively. (b) 2D map of the electric field intensity in log scale where (b) NP thickness = 1 nm (c) NP thickness = 9 nm. Wavelength for the 2D map was set to lower polariton peak position. The illumination is set along 65° degree. The electric field intensity is plotted on a log scale.

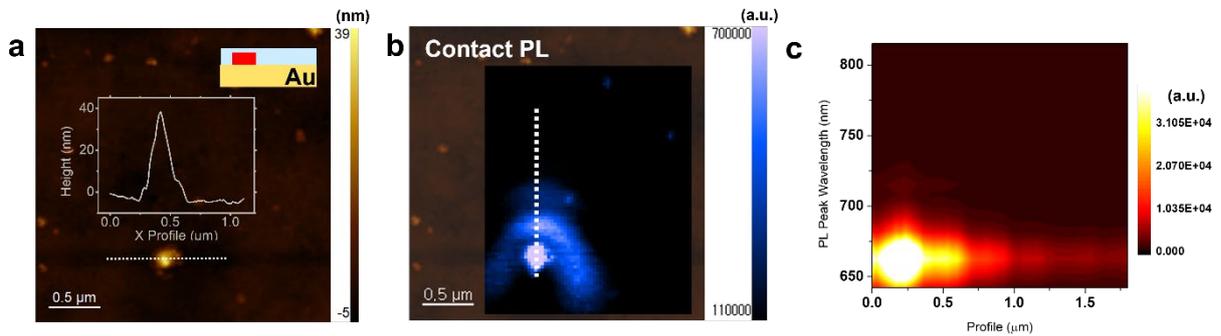

Figure S3. (a) Topography image of $Al_2O_3$ (5 nm)/NP/Au and (b) contact mode TEPL at 664 nm (c) TEPL intensity Profile along dotted line in (b) with respect to PL peak wavelength. Inset in (a) indicates the height profile along the dotted line in (a).

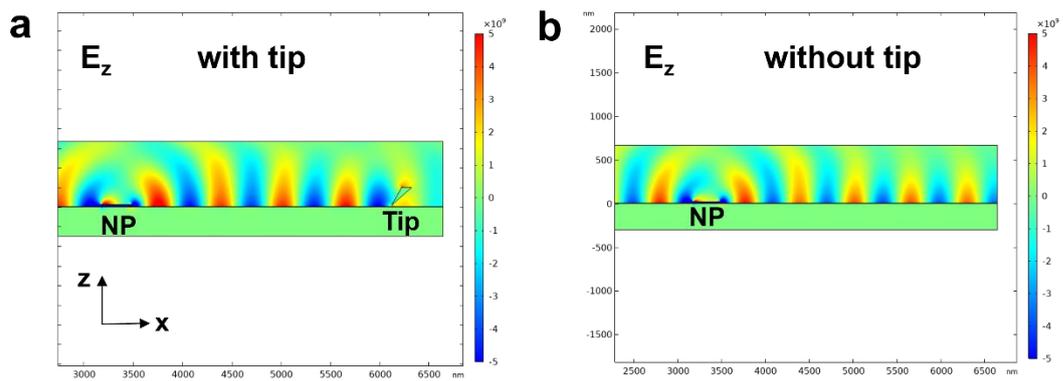

Figure S4. (a, b) $E_z$ field map of TM polarized NP emission on Au (a) with and (b) without the tip.

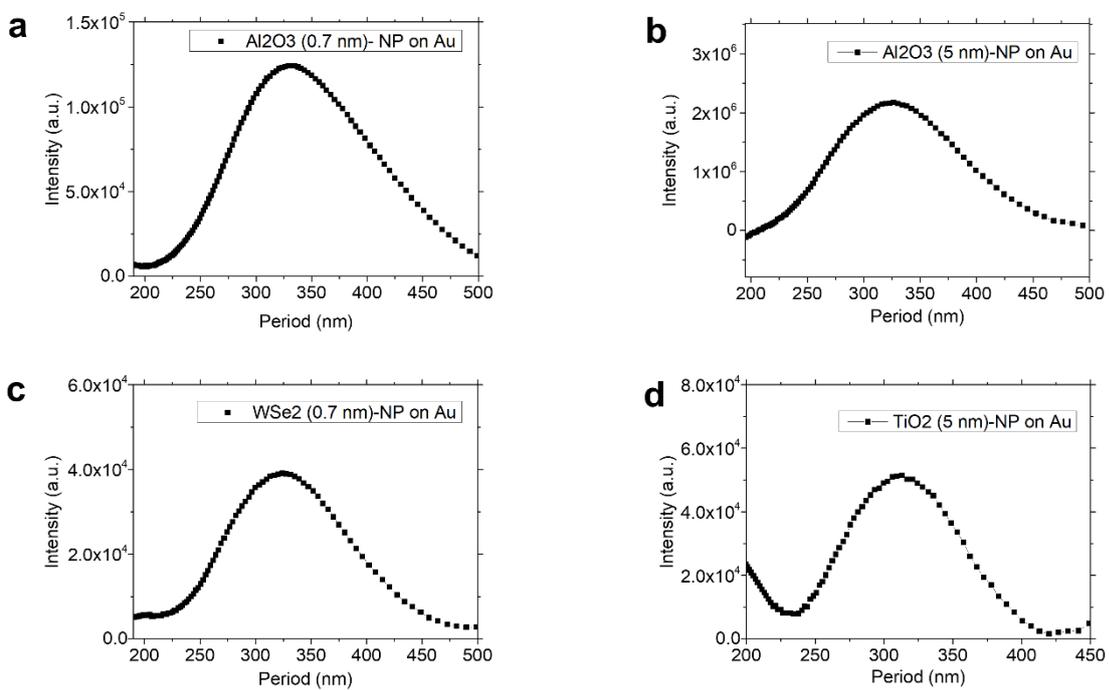

Figure S5. Fourier transforms extracted from the near field TEPL maps for (a) $Al_2O_3$ (0.7 nm)/NP/Au (b) $Al_2O_3$ (5 nm)/NP/Au (c) $WSe_2$ (0.7 nm)/NP on Au and (d) $TiO_2$ (5nm)/NP/Au.

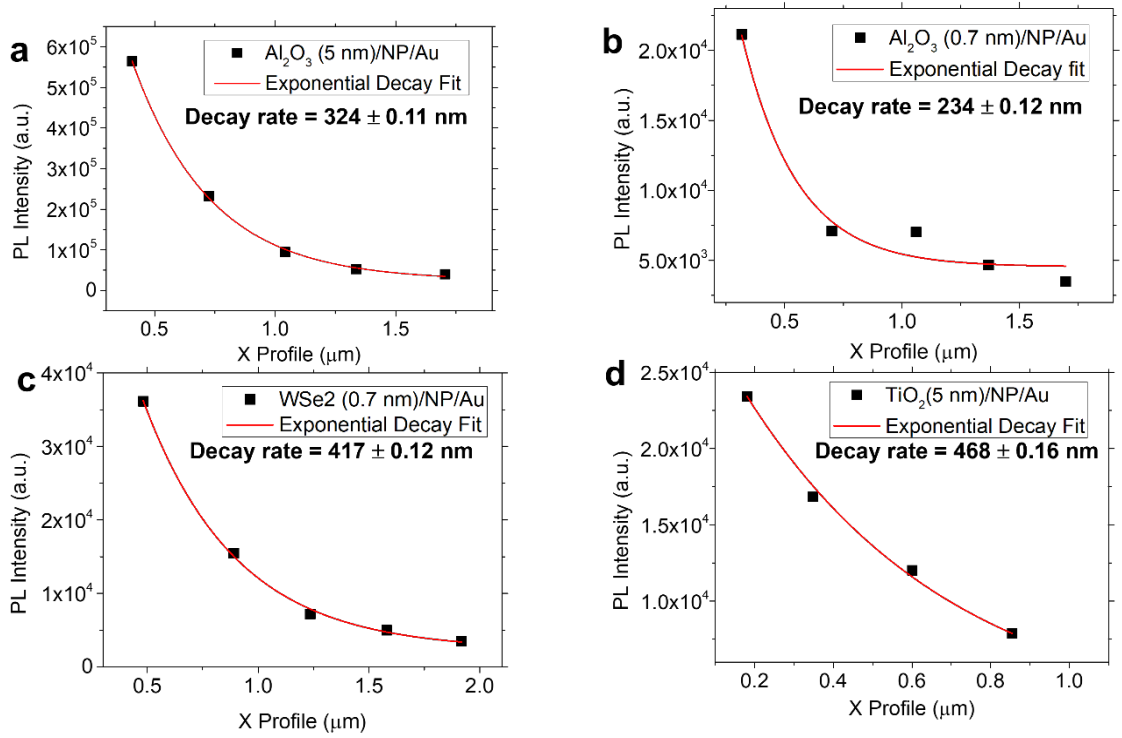

Figure S6. Tapping mode TEPL intensity at antinodes of the fringes of (a) $Al_2O_3$ (5 nm)/NP/Au (b) $Al_2O_3$ (0.7 nm)/NP/Au (c) $WSe_2$ (0.7 nm)/NP/Au (d) $TiO_2$ (5 nm)/NP/Au.

Decay constants ($\beta$) were obtained by fitting exponential decay curve as follows.

$$y = y_o + A_o e^{-\frac{x-x_o}{\beta}}$$

The exponential decay fittings were iterated 500 times with a tolerance of $1*10^{-15}$.

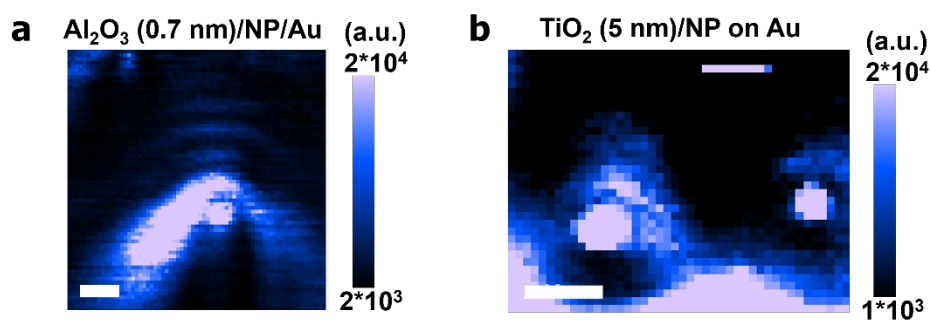

Figure S7. Hyperspectral TEPL intensity map (λ=664 nm) of (a) Al$_2$O$_3$ (5 nm)/NP/Au and (b) TiO$_2$ (5 nm)/NP/Au.

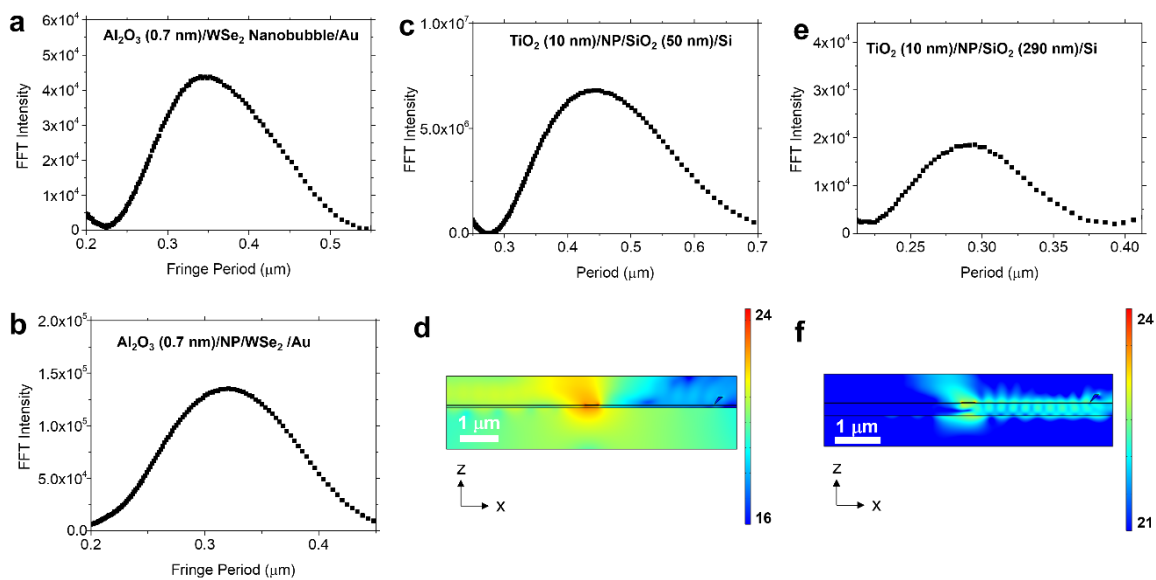

Figure S8. Fourier transformed periodic fringe profiles extracted from the near field TEPL maps for (a) Al$_2$O$_3$ (0.7 nm)/WSe$_2$ nanobubbles/Au and (b) Al$_2$O$_3$ (5 nm)/NP/WSe$_2$/Au. (c, d, e, f) Fourier transformed periodic fringe profiles extracted from the near field TEPL maps for (c) TiO$_2$ (10 nm)/NP/SiO$_2$ (50 nm)/Si and (e) TiO$_2$ (10 nm)/NP/SiO$_2$ (290 nm)/Si. Simulated E-field strength map of (d) TiO$_2$ (10 nm)/NP/SiO$_2$ (50 nm)/Si and (f) TiO$_2$ (10 nm)/NP/SiO$_2$ (290 nm)/Si.

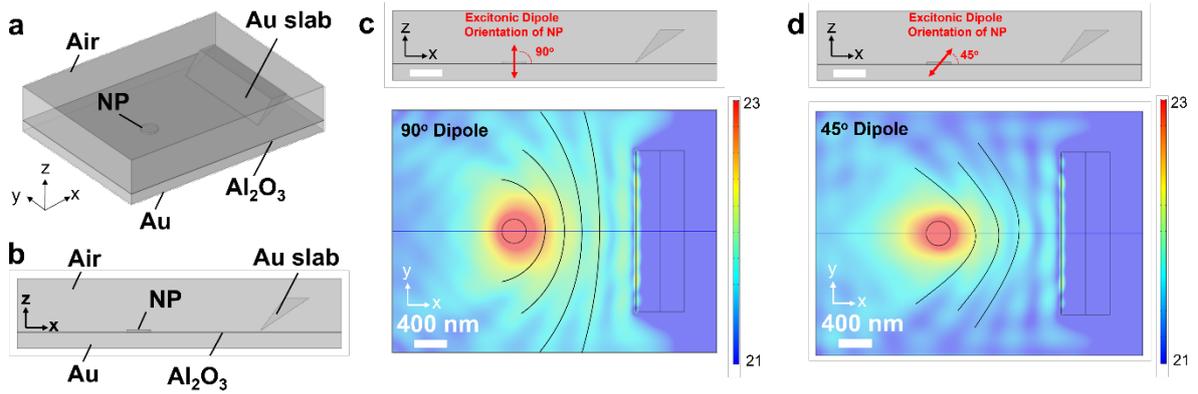

Figure S9. (a, b) Schematic representation of Al$_2$O$_3$ (5 nm)/NP/Au with Au Slab with (a) isometric view and (b) side view used for the 3D simulations. Au slab represents the trajectory of Au tip during the scan. (c, d) Side view with excitonic dipole orientation (top) and in-plane E-field strength map at height = 5 nm above the dielectric layer (bottom) with (c) 90º (z-directional) excitonic dipole (d) 45º angle of excitonic dipole in XZ plane with respect to the sample plane. Scale bars indicate 400 nm.

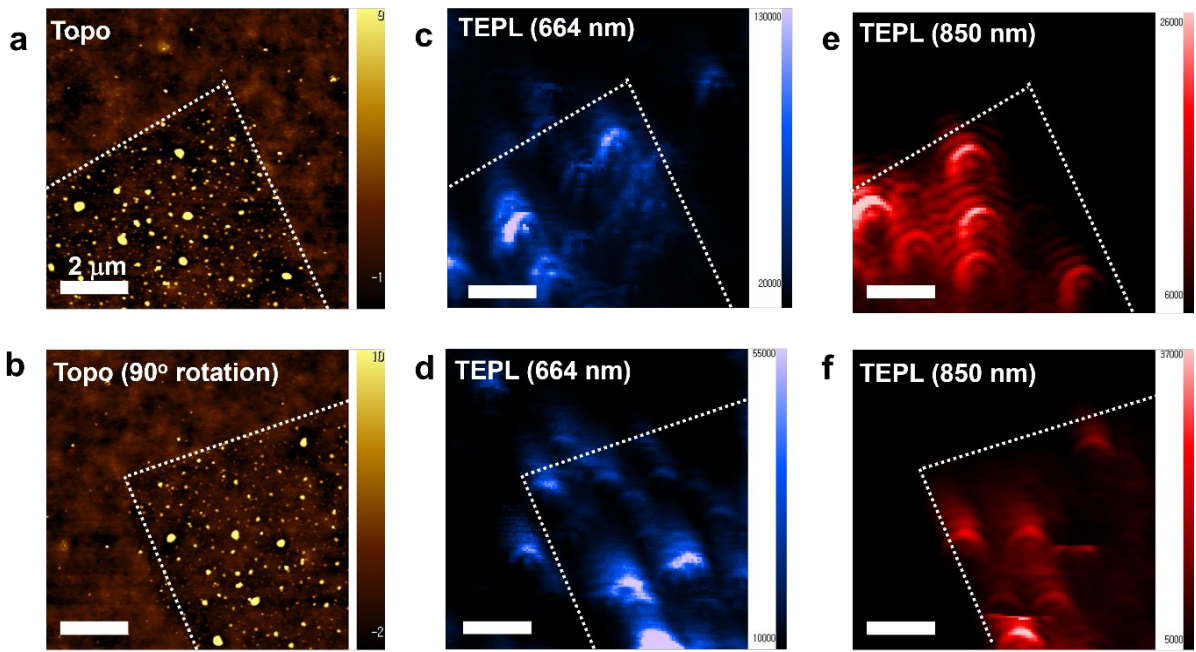

Figure S10. (a, b) Topography image of Al$_2$O$_3$/NP/WSe$_2$/Au (a) before and (b) after the 90º rotation. (c, d) The emission from the NP (c) before and (d) after the 90º rotation. (e, f) The emission from the WSe$_2$ nanobubbles (c) before and (d) after the 90º rotation. The fringe direction does not change upon rotation of sample as seen in (c-f) Scale bars indicate 2 μm.

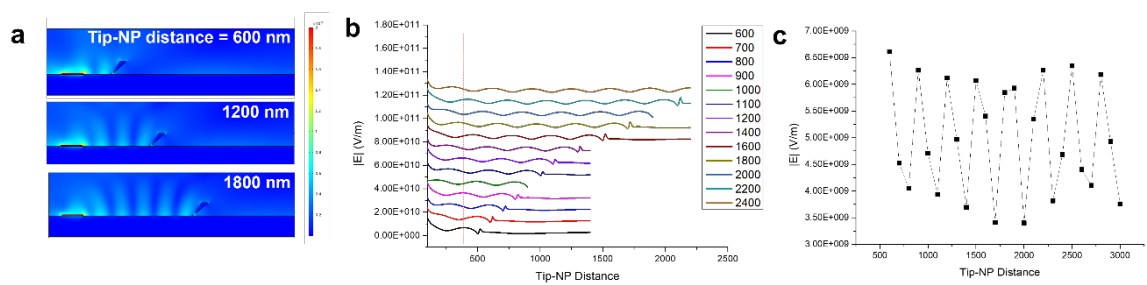

Figure S11. (a) Simulated 2D map of E field fringes between NP and tip with respect to tip-NP distance. (b) Extracted E-field strength profile at z= 10 nm as a function of distance. (c) E-field strength at the fixed point where 400 nm away from NP (Red dotted line).

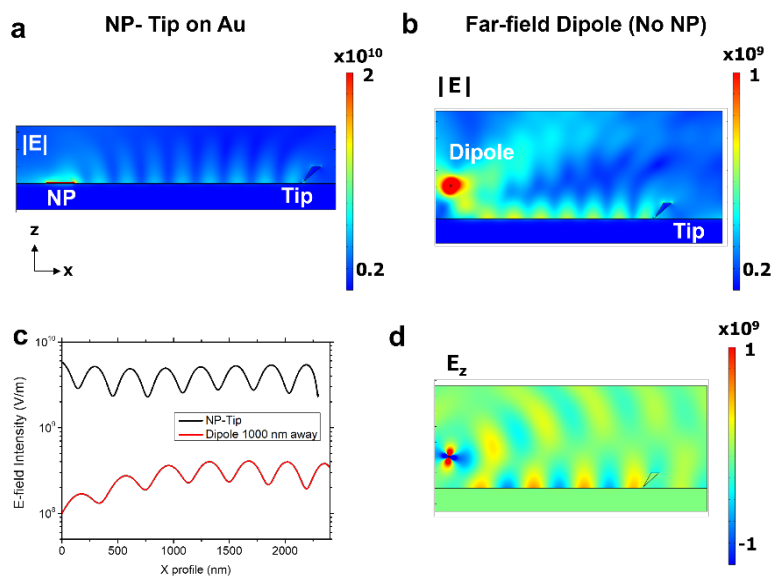

Figure S12. (a, b) E-field strength map of (a) Al$_2$O$_3$ (5 nm)/NP/Au with tip system and (b) Al$_2$O$_3$ (5 nm)/Au with tip and dipole 1000 nm away from the surface without NP. (c) Comparison of E-field intensity of (a) and (b) at height = 10 nm (d) E$_z$ field profile of Al$_2$O$_3$ (5 nm)/Au with tip and dipole 1000 nm away from the surface without NP.

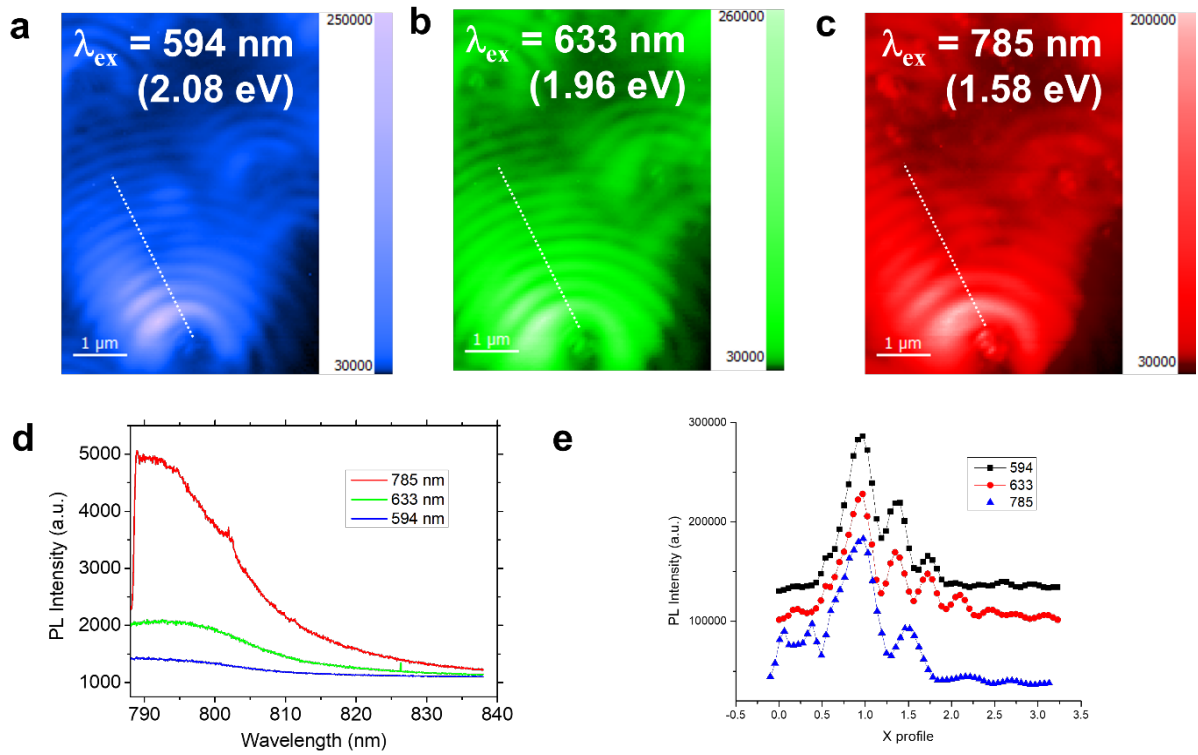

Figure S13. (a-c) Fringe from WSe2 nanobubbles emitting PL at 800 nm under (a) 594 nm (b) 633 nm (c) 785 nm excitation laser. (d) Emission spectrum of WSe$_2$ nanobubbles and (e) Fringe profile along with dotted line.

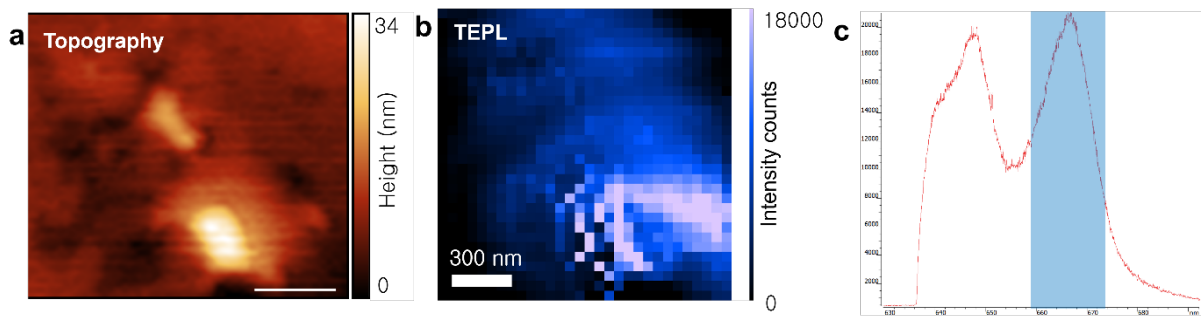

Figure S14. (a) Topography (b) Contact mode TEPL map and (c) corresponding spectrum of WS$_2$ nanobubble.